\documentclass[a4paper,fleqn,usenatbib]{mnras}
\usepackage{color}
\usepackage{graphicx,url,amssymb,amsmath,longtable,rotating,units,wasysym,listings,bm}
\usepackage{amsmath}	
\usepackage{amssymb}	
\usepackage{mathptmx}
\usepackage{ulem}

\newcommand\simlt{\lower.5ex\hbox{$\; \buildrel < \over \sim \;$}}
\newcommand\simgt{\lower.5ex\hbox{$\; \buildrel > \over \sim \;$}}

\DeclareMathOperator\arctanh{arctanh}

\title{The spectrum of a fast shock breakout from a stellar wind}
\author[K. Ioka et al.]{Kunihito Ioka$^1$, Amir Levinson$^{1,2}$ \& Ehud Nakar$^2$
\\
$1$Center for Gravitational Physics, Yukawa Institute for Theoretical Physics, Kyoto University, Kyoto 606-8502, Japan\\
$2$ The Raymond and Beverly Sackler School of Physics and Astronomy, Tel Aviv University, Tel Aviv 69978, Israel\\
}

\begin{document}
\label{firstpage}
\pagerange{\pageref{firstpage}--\pageref{lastpage}}
\maketitle
	
\begin{abstract}
  The breakout of a fast  ($>0.1 c$), yet sub-relativistic shock from a thick stellar wind
  is expected to produce a pulse of X-rays with a rise time of seconds to hours.
  Here, we construct a semi-analytic model for the breakout of a sub-relativistic,
  radiation-mediated shock from a thick stellar wind,
  and use it to compute the spectrum of the breakout emission.
  The model incorporates photon escape through the finite optical depth wind,
  assuming a diffusion approximation and
  a quasi-steady evolution of the shock structure during the breakout phase.
  We find that in sufficiently fast shocks,
  for which the breakout velocity exceeds about $0.1c$,
  the time-integrated spectrum of the breakout pulse is non-thermal,
  and the time-resolved temperature is expected to exhibit substantial decrease
  (roughly by one order of magnitude) during breakout,
  when the flux is still rising,
  because of the photon generation by the shock compression associated with the photon escape.
  We also derive a closure relation between the breakout duration,
  peak luminosity, and characteristic temperature that can be used to test
  whether an observed X-ray flare is consistent with being associated with
  a sub-relativistic shock breakout from a thick stellar wind or not.
  We also discuss implications of the spectral softening
  for a possible breakout event XRT 080109/SN 2008D.
\end{abstract} 
%


\section{Introduction}
Radiation-mediated shocks (RMS) play a key role in shaping the early emission
observed in various types of cosmic explosions. 
The radiation trapped inside the shock is released upon breakout of the shock
from the thick envelope enshrouding the source of the explosion.
The structure and velocity of the shock,
and the characteristics of the consequent emission depend on the type of the progenitor,
the explosion energy, and the angular extent of the ejecta
\citep[for a recent review see][]{waxman2017}.
While in certain situations the shock created in the explosion may become
relativistic \citep{Tan2001,nakar2012,Kyutoku2014}, in the majority of the events
it is sub-relativistic, albeit fast.

In progenitors which are surrounded by sufficiently tenuous circumstellar medium,
the breakout occurs as the shock approaches the sharp edge of the stellar envelope,
whereupon it undergoes an abrupt transition from an RMS to a collisionless shock.
However, if the progenitor is surrounded by a thick enough stellar wind,
the shock continues to be radiation mediated also
after it emerges from the stellar envelope and continues to propagate down the wind
\citep{Campana2006,Waxman2007,Soderberg2008,katz2010,Balberg2011}.
The shock physical width then increases during its propagation
since the optical depth of the wind decreases.

In relativistic shocks, the breakout is significantly delayed,
owing to opacity self-generation. 
It has been shown \citep{granot2018} that for a sufficiently high shock Lorentz factor
$\gamma_s$, at which the immediate downstream temperature approaches $m_e c^2/3$, 
the transition to a collisionless shock occurs at a radius
beyond which the (pair unloaded) Thomson optical depth to infinity ahead of the shock is
$\tau\simeq m_e \gamma_s /m_p$.
This is because beneath this radius, the number of escaping photons
that are backscattered into the flow direction 
is larger than the number of electrons in the far upstream flow,
giving rise to an accelerated pair production.
In highly non-relativistic RMS, where pair production is negligible,
the transition takes place once the optical depth 
ahead of the shock approaches $c/v_s$, where $v_s$ is the shock velocity.
One might naively suspect that 
in the intermediate regime, specifically for shock velocities
$v_s/c > \sqrt{m_e/m_p}$, pair production may become substantial, 
owing to a steepening of the shock or
formation of a subshock that enhances the temperature.
If true, it can lead to a delayed breakout as in the relativistic case.
However, this can only happen if photon generation within a 
diffusion length is not efficient enough during the breakout.
We show below that the temperature does not rise,
  or rather decreases, during the photon escape.
In any event, once breakout commences and the radiative losses start increasing, 
the shock structure gradually changes,
and this might affect the downstream temperature and the spectrum of 
emitted radiation.
Thus, detailed calculations of the temperature profile during the breakout phase are
desirable in order to address these issues.    
Such calculations can be performed using numerical methods,
like the one developed by \cite{ito2018a}.
However, until this method will be adjusted to non-relativistic shocks,
one may resort to an approximate analytic approach in order to get physical insight.

We note that a sub-relativistic shock breakout from a wind is expected
for a smaller progenitor than a red supergiant star.
Such a progenitor like a Wolf--Rayet (WR) star
is known to eject winds before exploding as a Type Ic or Ib supernova
\citep[e.g.,][]{Tanaka2009,Gal-Yam2014}.
Although the shock breakout from a wind is recently observed to be common
in Type II supernovae
\citep[e.g.,][]{Moriya2011,Ofek2014,Yaron2017,Forster2018},
the candidates for sub-relativistic breakouts are still rare
\citep{Soderberg2008,Mazzali2008,Modjaz2009}
because of its short, faint, and X-ray signal.
Therefore, theoretical predictions are important
for maximizing the observational prospects.
  A fast shock breakout may also happen in
  a binary neutron star merger like GW170817 \citep{GW170817,EM17}
  when a cocoon breaks out from the merger ejecta,
  and the breakout emission can dominate the jet emission
  \citep{Kasliwal+17,Gottlieb+18,Nakar+18}
  because an off-axis jet is faint for a large viewing angle \citep{IN18}.

At sufficiently low shock velocities, $v_s/c\ll1$,
the structure of an RMS can be computed analytically 
by employing the diffusion approximation.
Such an approach has been undertaken by, e.g.,
\cite{weaver1976,blandford1981a,blandford1981b,katz2010}.
While \cite{blandford1981b} considered shocks in which photon advection by the upstream
flow dominates over photon production,
that might be suitable for gamma-ray burst outflows,
\cite{weaver1976} and \cite{katz2010} computed 
the structure of the RMS under conditions  more suitable for shock breakout in stellar explosions.    
These studies indicate that the dominant photon source in such shocks is bremsstrahlung 
emission by the shocked electrons,
and that once the shock velocity exceeds about $0.05 c$ the radiation in the immediate 
shock downstream falls out of thermodynamic equilibrium and
its temperature becomes extremely sensitive to the shock velocity: $T_d \propto v_s^8$.
This strong dependence of temperature on velocity 
has important implications for the observational diagnostics of the breakout signal. 

The analyses outlined above assume that the shock is infinite.
However, as mentioned earlier, during the breakout phase 
an increasing fraction of the shock energy is radiated away,
and the question arises as to what effect this might have on the 
structure and emission of the shock.
Attempts to address this issue in case of a sudden breakout from a stellar envelope 
have been made recently using time-dependent models \citep{sapir2011,katz2012,Sapir2013}.
Here, we consider a gradual breakout from a stellar wind. 
Under the assumption that the shock continuously adjusts to local conditions,
so that it can be considered quasi-steady at
any given time, we construct an analytic model that takes into account photon escape,
and compute the temperature profile inside the
shock for different values of the energy fraction escaping the system,
by solving the photon transfer equation in the diffusion limit. 
The resultant temperature profiles are shown to be insensitive to the closure condition
(e.g., angular distribution of the radiation) invoked upstream of the shock. 
We find that in fast shocks ($v/c\simgt0.1$) the peak temperature decreases
as the energy fraction escaping the shock increases.
We also discuss the implications for the observed breakout signal,
in particular to compare with a possible breakout event XRT 080109/SN 2008D.


\section{Analytic model of the shock structure}
Consider a sub-relativistic RMS propagating in the negative $x$-direction
(${\bf v}= v\hat{x}$ in the shock frame).
We suppose that in the frame of the shock the flow is stationary ($\partial_t=0$),
and choose $x=0$ to be the boundary
upstream from which photons escape to negative infinity,
where the coordinate $x$ is measured in the shock frame.  
We further assume that the pressure is dominated by the radiation everywhere, and 
neglect the plasma pressure.
In what follows,
upstream quantities are labelled with a subscript $u$,
whereby $v_u$, $n_u$, $n_{\gamma u}$, and $p_{\gamma u}$ 
are the velocity, plasma density, photon density, and radiation pressure in the far upstream flow,
as measured in the shock frame.  
As will be confirmed,
for the range of shock velocities considered below pair production can be ignored,
hence the plasma density satisfies $n_p=n_e \equiv n$ everywhere.   
%

%

In the diffusion limit, the photon number flux and radiation energy flux are given,
to second order in $v_u/c$, by 
${\bf j}_\gamma =j_\gamma \hat{x}$ and ${\bf f}_\gamma = f_\gamma\hat{x}$, respectively, with
\begin{eqnarray}
j_\gamma &=& v n_\gamma - \frac{c}{3n\sigma_T}\frac{d n_\gamma}{dx} ,\label{eq:j_gamma} \\
f_\gamma &=& 4 p_\gamma v -\frac{c}{n\sigma_T} \frac{d p_\gamma}{d x},
\end{eqnarray}
where $\sigma_T$ is the Thomson cross section \citep{blandford1981a,blandford1981b}.
The fluid equations in the shock frame are reduced to:
\begin{eqnarray}
m_p n v= m_p n_u v_u &\equiv & J,\label{eq:shc-1}\\
\frac{d}{dx}(J v + p_\gamma) &=& 0,\label{eq:shc-2}\\
\frac{d}{dx}\left(J v^2/2 + f_\gamma \right) &=& 0.\label{eq:shc-3}
\end{eqnarray}
The above equations can be rendered dimensionless upon defining
$\tilde{p}_\gamma=p_\gamma/Jv_u$, $\tilde{v}=v/v_u$, and $d\tau^\star= (v_u/c) n_e\sigma_T dx$
(note that the velocity is included in the definition of the optical depth). 
Integrating Eqs.~(\ref{eq:shc-2}) and (\ref{eq:shc-3}), using $\tilde{v}_u=1$, gives:
\begin{eqnarray}
  \tilde{v}+\tilde{p}_\gamma &=& 1+\tilde{p}_{\gamma u},
  \label{eq:conti}\\
\frac{d\tilde{p}_\gamma}{d\tau^\star} &=& -\frac{1}{2} + \frac{1}{2}{\tilde v}^2 +4\tilde{p}_\gamma {\tilde v}-  \frac{1}{2}\tilde{f}_{\gamma u},
\label{eq:EOM}
\end{eqnarray}
where
\begin{eqnarray}
  \tilde{f}_{\gamma u} = \frac{2f_{\gamma u}}{J v_u^2}
  \label{eq:tildef}
\end{eqnarray}
denotes the fraction of shock energy that escapes to infinity, and
must satisfy $\tilde{f}_{\gamma u}>-1$.
Equations (\ref{eq:conti}) and (\ref{eq:EOM}) admit an analytic solution
for arbitrary values of $\tilde{p}_{\gamma u}$ and $\tilde{f}_{\gamma u}$:
\begin{eqnarray}
  \tilde{v}=\frac{4}{7}(1+\tilde{p}_{\gamma u}) +\frac{\eta}{7}
  \tanh\left[\frac{\eta}{2}(\tau_0-\tau^\star)\right],
\label{eq:v_prof}
\end{eqnarray}
where
\begin{eqnarray}
  \eta = \sqrt{(3-4\tilde{p}_{\gamma u})^2 + 56 \tilde{p}_{\gamma u}-7\tilde{f}_{\gamma u} }
  \label{eq:defeta}
\end{eqnarray}
and 
\begin{eqnarray}
  \tau_0=\frac{2}{\eta} \arctanh \left[\frac{3-4\tilde{p}_{\gamma u}}{\eta}\right]
  =\frac{1}{\eta}\ln \frac{\eta+3-4\tilde{p}_{\gamma u}}{\eta-3+4\tilde{p}_{\gamma u}},
\label{eq:tau0}
\end{eqnarray}
for the boundary condition $\tilde{v}_u=1$ at $\tau^\star=0$.
Here, $\tau_0$ is roughly the center of the shock transition layer 
(the precise location is slightly shifted, depending on the choice of parameters).
It can be readily verified that the downstream velocity,
$\tilde{v}_d = \tilde{v}(\tau^\star\rightarrow \infty)$,
satisfies $\tilde{v}_d \ge 0$ for $\tilde{f}_{\gamma u} \ge -1$, and that
no physical solutions exist for $\tilde{f}_{\gamma u} < -1$, as expected.
It is also seen that with the choice $(d\tilde{p}_{\gamma}/d\tau^\star)_u = 0$,
or $\tilde{f}_{\gamma u}=8\tilde{p}_{\gamma u}$  in our notation,
which corresponds to an infinite shock with no escape,
the solution obtained by \cite{blandford1981b} is recovered upon 
defining $\tau^\star - \tau_0 \rightarrow \tau^\star$.  

\subsection{A note on boundary  conditions}
The solution described by Eq. (\ref{eq:v_prof}) depends, formally,
on two free parameters, $\tilde{p}_{\gamma u}$ and $\tilde{f}_{\gamma u}$.
In an infinite shock with a cold upstream,
the physical choice is $\tilde{p}_{\gamma u} = \tilde{f}_{\gamma u} = 0$, since 
photons cannot reach distances larger than a few diffusion lengths upstream of the shock.
In a finite shock with photon escape,
the value of $\tilde{f}_{\gamma u}$ specifies the energy fraction which is radiated away.
However, within the diffusion 
approximation the value of $\tilde{p}_{\gamma u}$ is uncertain,
and there seems to be a degeneracy in the solution.
In reality
$\tilde{p}_{\gamma u}$ is fixed by additional physics beyond the diffusion approximation.
One naively anticipates some relation of
the form $f_{\gamma u} = e_{\gamma u} v_{rad}$, where $e_{\gamma u}$ is
the energy density of the escaping radiation 
and $v_{rad}$ is some effective velocity
that depends on the angular distribution of the radiation and, perhaps, some other details. 
If, for instance, one invokes complete beaming at the boundary $x=0$,
then $e_{\gamma u}= p_{\gamma u}$, $v_{rad}=-c$, and 
$f_{\gamma u} =-p_{\gamma u} c$.
In the other extreme, if the escaping radiation is taken to be sufficiently isotropic
just upstream of the shock, then the effective velocity is 
$v_{rad}\simeq -v_u$, the equation of state is $e_{\gamma u}= 3 p_{\gamma u}$,
and thus $f_{\gamma u} =-3 p_{\gamma u} v_u$.
In reality a similar relation likely holds, but with a somewhat different prefactor.  
We shall henceforth adopt the relation
\begin{eqnarray}
  \tilde{f}_{\gamma u} =-2 \alpha \tilde{p}_{\gamma u},
  \label{eq:alpha}
\end{eqnarray}
and explore the dependence of the 
solution on the dimensionless parameter $\alpha$
(where a factor of $2$ comes from the normalization in Eq.~(\ref{eq:tildef})).
Note that $\alpha=c/v_u$ for the  complete beaming case and $\alpha=3$ for the isotropic case.
To check the sensitivity of the solutions to the choice of $\alpha$,
we examine a broader range of values, $1\le \alpha \le c/v_u$. 
In the above prescription $\alpha=1$ corresponds to a diffusion velocity $v_{rad}=-v_u/3$.

\subsection{Shock solutions}
\begin{figure}
\centering
\includegraphics[width=1\columnwidth]{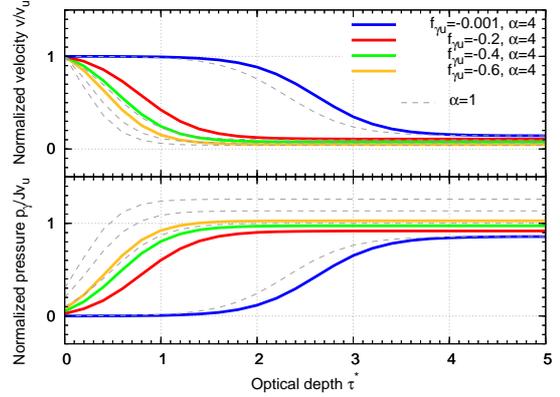}
\caption{Normalized velocity ({\it upper panel}) and pressure ({\it lower panel}) profiles
  plotted as functions of the optical depth $\tau^\star$
  for different values of the radiation energy flux upstream,
  $\tilde{f}_u=-0.001, -0.2, -0.4, -0.6$, 
  and $\alpha=4$ (complete beaming for $c/v_u=4$)
  and $\alpha=1$,
  where $\tilde{f}_{\gamma u} = -2\alpha\tilde{p}_{\gamma u}$.
  The blue line corresponds to an almost infinite shock.}
\label{fig:v&p}
\end{figure}

Solutions for the shock profile are exhibited in Fig.~\ref{fig:v&p}
for $v_u/c=0.25$ and different values of $\tilde{f}_{\gamma u}$ and $\alpha$.
As seen, photon escape leads to shock steepening, as expected.
Specifically, for given values of $v_u$ and $n_u$,
the shock transition layer becomes somewhat narrower as $\tilde{f}_{\gamma u}$ increases,
and the far downstream velocity $v_d$ becomes smaller, 
giving rise to correspondingly larger values of the downstream density $n_d$,
and somewhat larger values of the pressure $p_{\gamma d}$. 
Quite generally, the dependence of the solution on $\alpha$ is found to be rather weak
(as long as the shock velocity is sub-relativistic i.e., $c/v_u\simlt 3$).

\begin{figure}
\centering
\includegraphics[width=0.9\columnwidth]{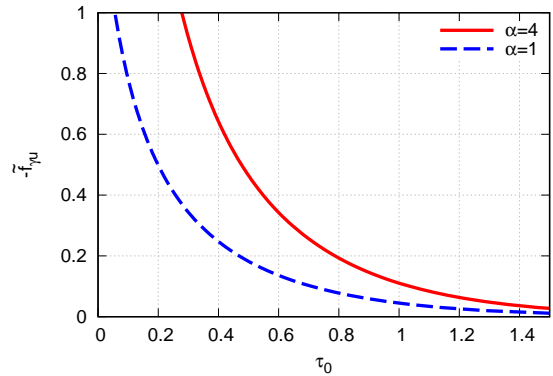}
\caption{A plot of $|{\tilde f}_{\gamma u}|$ versus $\tau_0$ for $\alpha=1$ (dashed line)
  and $\alpha=4$ (solid line).}
\label{fig:tau0-f}
\end{figure}

It is worth pointing out that, formally,
as the fraction of shock energy which is radiated away approaches unity, namely 
$\tilde{f}_{\gamma u} = -1$, we have $\eta =4+2/\alpha$ 
compared with $\eta=3$ in the infinite case ($\tilde{f}_{\gamma u} = 0$)
with Eqs.~(\ref{eq:defeta}) and (\ref{eq:alpha}).
Since the width of the transition layer scales as $\sim 2/\eta$
(see Eq. \ref{eq:v_prof}), it means that even 
in the presence of substantial losses the shock width remains of order $c/v_u$, and is reduced
only by a numerical factor, roughly $(4+2/\alpha)/3$.
This trend is seen in Fig.~\ref{fig:v&p}, and more clearly in Fig.~\ref{fig:tau0-f}, where
$|f_{\gamma u}|$ is plotted against $\tau_0$.
Fig.~\ref{fig:tau0-f} confirms that radiative losses commence roughly when the optical depth 
from the center of the shock transition layer to infinity
is about $c/v_u$ and become nearly maximal when it equals $(c/v_u)\ln[7/(1+4/\alpha)]/(4+2/\alpha)$,
approximately $0.3 c/v_u$ for $\alpha=4$.
The prime reason, as can be seen from Eq. (\ref{eq:alpha}) and Fig.~\ref{fig:v&p},
is that the upstream pressure required for 
substantial losses is a small fraction of the downstream pressure (unless $\alpha$ is small).
Even for $\alpha=1$ we find  $p_u/p_d\simeq0.2$ 
at $\tilde{f}_{\gamma u}=-0.5$, and a shock width of $\Delta \tau\simeq 0.5 c/v_u>1$.
It practically means that the diffusion approximation 
is a reasonable approximation even at large losses. 
On the other hand, the downstream velocity approaches zero as
$\tilde{f}_{\gamma u} \rightarrow -1$, implying increasingly strong compression
of the downstream layer as breakout proceeds and $|f_{\gamma u}|$ increases. 

\begin{figure*}
  \centering
\includegraphics[width=8 cm]{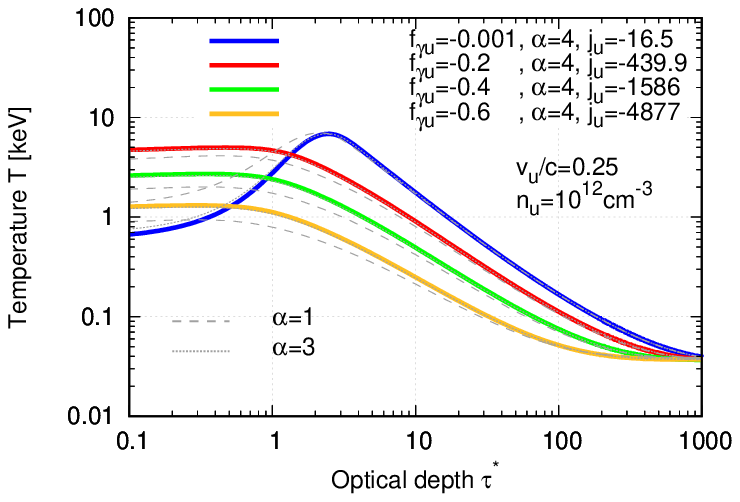} \includegraphics[width=8 cm]{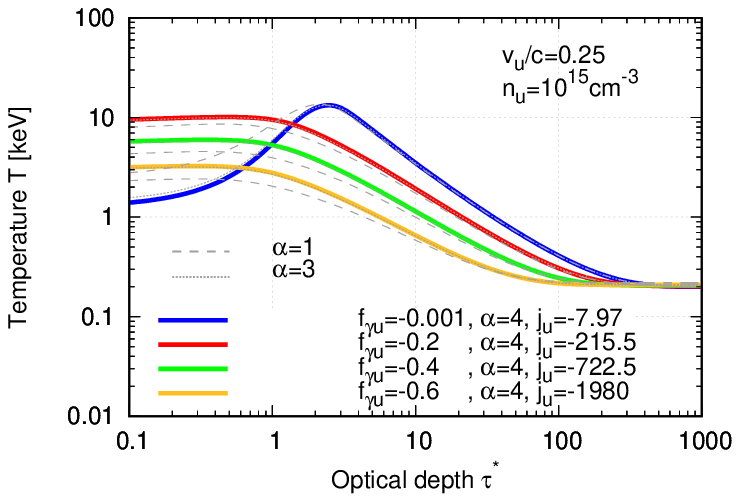}
\caption{Temperature profiles as functions of the optical depth $\tau^\star$
  for the upstream velocity $v_u/c=0.25$ and densities
  $n_u=10^{12}$ cm$^{-3}$ (left-hand panel) and $n_u=10^{15}$ cm$^{-3}$ (right-hand panel).
  We take the radiation energy flux upstream
  $\tilde{f}_u=-0.001, -0.2, -0.4, -0.6$,
  and $\alpha=4$ (complete beaming for $c/v_u=4$; {\it solid lines}),
  $\alpha=3$ (isotropic case; {\it dotted lines}), and $\alpha=1$ ({\it dashed lines})
  where $\tilde{f}_{\gamma u} = -2\alpha\tilde{p}_{\gamma u}$.
  The photon number flux upstream is adjusted to satisfy the boundary condition
  $j_{\gamma u}=f_{\gamma u}/3kT_u$.
}
\label{fig:T}
\end{figure*}

\subsection{Computing the temperature profile}
The local temperature can be obtained from the relation
\begin{eqnarray}
  kT(x)=\frac{p_\gamma(x)}{n_\gamma(x)},
  \label{eq:T}
\end{eqnarray}
once the photon density is known.
The evolution of the latter is governed by the equation 
\begin{eqnarray}
\frac{d j_\gamma}{dx}= \dot{n}_\gamma,
\label{eq:dj_gamma/dx}
\end{eqnarray}
with $j_\gamma$ given by Eq. (\ref{eq:j_gamma}), 
where $\dot{n}_\gamma$ is a photon source that accounts for all emission and absorption processes.
Under the conditions envisaged here,
photon generation is dominated by bremsstrahlung emission of the hot electrons. 
Absorption can be accounted for by including the suppression factor
\begin{eqnarray}
  f_{ab}=1-\frac{n_\gamma}{n_{BB}}=1-2.4\times10^{-31}\Theta^{-3} n_\gamma,
  \label{eq:fab}
\end{eqnarray}
where $\Theta=kT/m_ec^2$ \citep{weaver1976}
that considerably simplifies the calculations and is sufficient for our purposes.
Specifically, $\dot{n}_\gamma = Q_{ff} f_{ab}$.
For the thermal bremsstrahlung source, we adopt the form (Eq.~(3.2) in \cite{weaver1976}):
\begin{eqnarray}
Q_{ff}=\frac{1}{2}\alpha_e n_p n_e \sigma_T c \Theta^{-1/2}\Lambda_{eff},
\label{eq:Q_ff}
\end{eqnarray}
expressed in terms of the fine structure constant $\alpha_e$ and the 
coefficient $\Lambda_{eff}=E_1(y) g_{eff}(y)$,
where $y \equiv h\nu_c/kT$,
$E_1(y)$ is the first-order exponential integral function,
which satisfies $E_1(y) \simeq -\ln y -0.5772$ at $y \ll 1$, and
$g_{eff} \approx 1.226-0.475 \ln y + 0.0013 (\ln y)^2$
is the Gaunt factor.
The cut-off frequency, $\nu_c(T,n)$, corresponds to the energy below which newly
generated soft photons are re-absorbed before being boosted to the thermal peak
by inverse Compton scattering.
It is derived in \cite{katz2010} and is given explicitly by their equation (11),
$y=({kT}/{m_e c^2})^{-9/4} ({\alpha_e g_{ff} (m_e c^2)^{-3} h^3 c^3 n}/{32\pi})^{1/2}$.
Combining equations (\ref{eq:j_gamma}), (\ref{eq:dj_gamma/dx}), and (\ref{eq:Q_ff}), we finally arrive at
\begin{eqnarray}
  \frac{d}{d\tau^\star}\left[\tilde{n}_\gamma \tilde{v} -\frac{1}{3}\frac{d\tilde{n}_\gamma}{d\tau^\star}\right]=\frac{\alpha_e}{2}\tilde{n}(c/v_u)^2\Theta^{-1/2} \Lambda_{eff}f_{ab}
  \equiv 3\tilde{v} \tilde{Q}_{\gamma},\label{eq:ph_genr-2}
\end{eqnarray}
with $\tilde{n}_\gamma\equiv n_{\gamma}/n_u=(m_p/m_e)(v_u/c)^2(\tilde{p}_\gamma/\Theta)$
for a given solution $\tilde{p}_\gamma (\tau^\star)$ of
the shock equations, specifically Eqs. (\ref{eq:conti}) and (\ref{eq:v_prof}).
Note that the photon flux is normalized as ${\tilde j}_{\gamma}\equiv j_{\gamma}/n_u v_u=
\tilde{n}_\gamma \tilde{v} -(1/3){d\tilde{n}_\gamma}/{d\tau^\star}$.

This equation is subject to the boundary conditions
$dj_\gamma/d\tau^\star \rightarrow 0$ at $\tau^\star\rightarrow \infty$, and 
\begin{eqnarray}
  j_{\gamma u} = \frac{f_{\gamma u}}{3kT_u}
  \label{eq:jgu}
\end{eqnarray}
at $\tau^\star =0$, where $T_u$ is the value of the temperature there.
The solution that satisfies the boundary conditions is obtained with the Green's function method as
\begin{eqnarray}
  {\tilde n_{\gamma}}({\tilde x})=
  \frac{{\tilde j_{\gamma u}}}{{\tilde v_{\rm eff}}({\tilde x})}
  + \int_{0}^{\infty} G({\tilde x}, {\tilde y})
  {\tilde Q_{\gamma}}({\tilde y}) d{\tilde y},
  \label{eq:nG}
\end{eqnarray}
where we introduce a new coordinate,
\begin{eqnarray}
  d{\tilde x}=3 (v_u/c) n_u \sigma_T dx = 3 \tilde{v} d\tau^\star,
\end{eqnarray}
and the Green's function is
\begin{eqnarray}
  G({\tilde x}, {\tilde y})=\left\{
  \begin{array}{ll}
    \frac{\displaystyle e^{{\tilde x}-{\tilde y}}}{\displaystyle {\tilde v_{\rm eff}}(\tilde y)},& ({\tilde x}\le {\tilde y})\\
    \frac{\displaystyle 1}{\displaystyle {\tilde v_{\rm eff}}({\tilde x})},& ({\tilde x}>{\tilde y})
  \end{array}\right.
\end{eqnarray}
with an effective velocity
\begin{eqnarray}
  \frac{1}{{\tilde v_{\rm eff}}({\tilde x})}
  =\int_{\tilde x}^{\infty} \frac{e^{-({\tilde x}'-{\tilde x})}}{{\tilde v}({\tilde x}')} d{\tilde x}'.
\end{eqnarray}
The boundary condition $dj_{\gamma}/d\tau^\star \to 0$
at the far downstream, $\tau^\star \to \infty$, is satisfied
thanks to the suppression factor $f_{ab}$ in Eq.~(\ref{eq:fab}),
that leads to a thermodynamic equilibrium, $\tilde{Q}_{\gamma} \to 0$.
A numerical solution of Eq.~(\ref{eq:nG}) is obtained through iteration:
$\tilde{Q}_{\gamma}$ is calculated from $\Theta$ (and $\tilde{n}_{\gamma}$ in $f_{ab}$)
with Eq.~(\ref{eq:ph_genr-2}),
$\tilde{n}_{\gamma}$ is calculated from $\tilde{Q}_{\gamma}$ with Eq.~(\ref{eq:nG}),
and $\Theta$ is calculated from $\tilde{n}_{\gamma}$ with Eq.~(\ref{eq:T}).
The convergence of the iteration becomes slow for large $|\tilde{f}_{\gamma u}|$.
The photon flux upstream, ${\tilde j}_{\gamma u}$, is adjusted
to satisfy the boundary condition in Eq.~(\ref{eq:jgu})
by the second numerical iteration.

Figure~\ref{fig:T} shows the temperature profiles thereby computed
for an upstream velocity $v_u/c=0.25$, two values of the density, $n_u=10^{12}$ cm$^{-3}$ (left-hand panel) 
and  $n_u=10^{15}$ cm$^{-3}$ (right-hand panel), and different values of the escape parameter,
$\tilde{f}_u=-0.001, -0.2, -0.4$, and $-0.6$, as indicated.
The thick solid lines are solutions obtained for 
$\alpha=4$ in Eq.~(\ref{eq:alpha}) (complete beaming),
the thin dotted lines for $\alpha=3$ (isotropic case),
and the dashed lines for $\alpha=1$.
The case $\tilde{f}_u=-0.001$, that corresponds to a nearly infinite shock,
is consistent with the previous calculations outlined in
\cite{weaver1976}\footnote{The temperature profile at small optical depths,
  below $\tau^\star\simeq 0.3$, is somewhat
  different from that obtained in \cite{weaver1976},
  owing to the different boundary condition used by this author.
  However, this does not affect the solution in the entire range.}
and \cite{katz2010} for $\alpha=4$ and $3$.
For $\alpha=1$, the drop-off at small optical depths somewhat deviates
from the infinite shock solution.
This is due to the higher upstream pressure required to obtain the same losses.
We verified that it does converge to the infinite 
shock case for smaller values of $\tilde{f}_{\gamma u}$.

\begin{figure}
  \centering
\includegraphics[width=8cm]{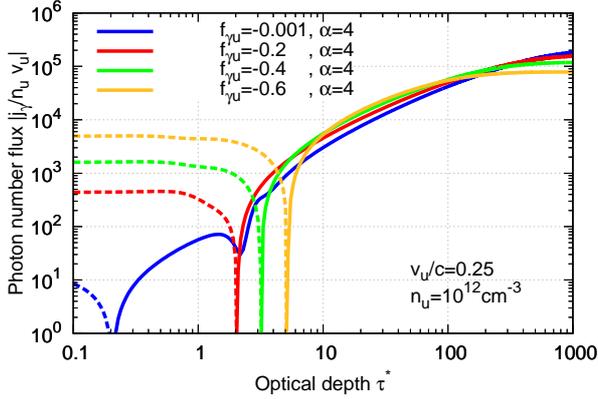}
\caption{Photon number fluxes $|{\tilde j}_{\gamma}|$ as functions of the optical depth $\tau^\star$
  for the upstream velocity $v_u/c=0.25$ and density $n_u=10^{12}$ cm$^{-3}$.
  The negative fluxes are shown by dashed lines.
  We take the radiation energy flux upstream
  $\tilde{f}_u=-0.001, -0.2, -0.4$, and $-0.6$,
  and $\alpha=4$ (complete beaming for $c/v_u=4$).
}
\label{fig:jgamma}
\end{figure}

Figure~\ref{fig:jgamma} shows the photon number fluxes $|{\tilde j}_{\gamma}|$
as functions of the optical depth $\tau^\star$
for the upstream velocity $v_u/c=0.25$ and density $n_u=10^{12}$ cm$^{-3}$
and different escape fractions $\tilde{f}_u=-0.001, -0.2, -0.4$, and $-0.6$ with $\alpha=4$.
The negative flux part is plotted by dashed lines.
As photons escape, the photon number fluxes change the sign at $\tau^{\star} \gtrsim 2$, i.e.,
after the shock compression as we can see from Fig.~\ref{fig:v&p}.
This is consistent with the picture that
the photons generated by the shock compression diffuse upstream
to be released as breakout emission.
Furthermore, we can check that
the sign changes at $\sim 1/4$ of the diffusion length $\Delta x \sim c/v n \sigma_T$.

\begin{figure}
  \centering
\includegraphics[width=8cm]{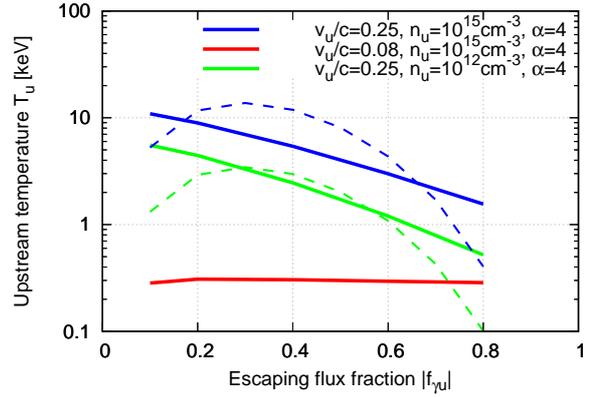}
\caption{Expected temperature evolution as a function of
  the radiation energy flux upstream (i.e., escaping flux fraction) $|\tilde{f}_u|$.
  The temperature peaks just after the breakout
  and exponentially decreases for fast shocks with $v_u/c \simgt 0.1$.
  We also plot the analytic approximation for the upstream temperature
  in Eq.~(\ref{eq:analytic}) by dashed lines
  with using $\Lambda_{eff}=20$ for $v_u/c=0.25$ and $n_u=10^{15}$ cm$^{-3}$,
  and $\Lambda_{eff}=40$ for $v_u/c=0.25$ and $n_u=10^{12}$ cm$^{-3}$.
}
\label{fig:Tevo}
\end{figure}

\begin{figure}
  \centering
\includegraphics[width=8cm]{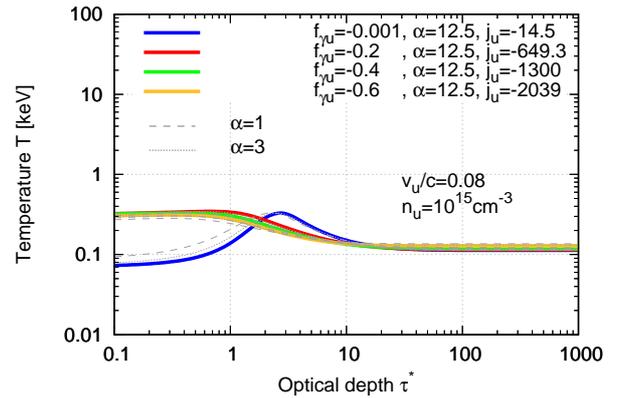}
\caption{Same as the right-hand panel in Fig. \ref{fig:T}, but for $v_u/c=0.08$.
}
\label{fig:T08}
\end{figure}

As Fig.~\ref{fig:T} indicates, the temperature of the escaping radiation decreases 
as the escaping flux $|{\tilde f}_{\gamma u}|$ increases.    This is seen more clearly 
in Fig~\ref{fig:Tevo}, where the upstream temperature is plotted against $|{\tilde f}_{\gamma u}|$.
The temperature declines exponentially as the loss rate increases.
Consequently, a spectral softening is anticipated during the course of the breakout.
The sole reason for this behavior is the increase in the shock compression ratio with
increasing losses, that leads to enhancement of the photon generation rate. 
To get some insight, we give a simple heuristic derivation of the upstream temperature:
the escaping photons is generated within $\sim 1/4$
of diffusion length $\Delta x \sim c/v n \sigma_T$
so as not to be swept downstream (see Fig.~\ref{fig:jgamma}).
From Eqs.~(\ref{eq:dj_gamma/dx}) and (\ref{eq:Q_ff}),
the photon number flux near the upstream boundary is approximately given by
\begin{eqnarray}
  j_{\gamma} \sim Q_{ff} f_{ab} \Delta x/4 \sim \alpha_e n c^2 \Theta^{-1/2} \Lambda_{eff}/8 v,
\end{eqnarray}
where we can put $f_{ab} \approx 1$.
Here, the photons are mainly generated after the deceleration down to $v \sim v_d$
(or the compression $n \propto 1/v$).
The number flux also satisfies the boundary condition in Eq.~(\ref{eq:jgu}).
By approximating $j_{\gamma} \sim |j_{\gamma u}|$ and $T \sim T_u$,
we obtain an analytic approximation for the upstream temperature as
\begin{eqnarray}
  kT_u \sim \frac{1}{m_ec^2}
  \left(\frac{4 m_p v_u^2 v_d^2 {\tilde f}_{\gamma u}}{3 \alpha_e c^2 \Lambda_{eff}}\right)^2.
  \label{eq:analytic}
\end{eqnarray}
This relation is shown in Fig.~\ref{fig:Tevo} by dashed lines,
and elucidates the dependence of $T_u$ on $v_d$,

To examine the dependence on the shock velocity,
we obtained solutions for $v_u/c=0.08$ (see Fig. \ref{fig:T08}).  
Contrary to the previous case,
the temperature profile is essentially independent of the loss rate.
This is expected since at such low velocity,
the characteristic length over which a full thermodynamic equilibrium is established
becomes comparable to the shock width.
The numerical result exhibited in Fig.~\ref{fig:T08} is
in a good agreement with the analytic result derived in 
equation (14) in \cite{katz2010}.

\subsection{Caveats} 
The quasi-steady, planar approximation invoked in our analysis is questionable, since at breakout the 
diffusion time across the shock and the expansion time are comparable.   On the one hand, dynamical effects might lead
 to some suppression of the photon production rate inside the shock and a corresponding increase of the immediate 
downstream temperature.  On the other hand, sphericity might give rise to adiabatic losses of the expanding shocked layer.  
What is the net effect on the observed temperature is difficult to assess.

Another concern is the validity of the RMS solution.  Within the diffusion approximation it has been found above that the radiation can 
support the shock even when the losses become large.  This requires causal contact across the shock to allow its 
adjustment of the changing conditions.   In reality a subshock will eventually form due to dynamical effects, which may alter the spectrum. 
Note, however, that the jump conditions across the entire shock transition are determined by the overall energy and momentum fluxes upstream
(i.e., incoming flux minus escaping flux). Consequently, the presence of a subshock will not affect the downstream temperature considerably.
It can lead to particle acceleration that might give rise to formation of a non-thermal tail via comptonization once the 
subshock energy becomes substantial, but is unlikely to alter significantly  the evolution of the downstream temperature during breakout.   
A complete treatment of these effects is beyond the scope of this paper
(see also Sec.~\ref{sec:LC}).

\section{Observational consequences}
As the shock emerges from the stellar envelope,
it starts propagating in the wind until breaking out at some
radius $R_{bo}$ at time $t_{bo}=R_{bo}/v_{bo}$ at velocity $v_{bo}$.
In the following, we adopt a wind profile of the form $\rho_w=A (r/R_\star)^{-2}$,
here $R_\star$ denotes the progenitor's radius.
The shock accelerates during propagating through the decreasing density profile
of the stellar envelope, and
the profile of the accelerated ejecta can be expressed in 
terms of the maximum velocity $v_0$ of the ejecta
subsequent to the shock emergence from the stellar envelope in the form \citep{nakar2010}
\begin{eqnarray}
E(v) = E_0(v/v_0)^{-\lambda} =\frac{4\pi c v_0}{\kappa}R_\star^2(v/v_0)^{-\lambda},
\end{eqnarray}
where $\lambda=(1+0.62 n)/0.19n$, and $1.5\le n\le3$
is the polytropic index that depends on the progenitor type.
Here, $E_0$ is the total energy contained in a shell
of optical depth $c/v_0$ with opacity $\kappa$.
The breakout velocity can be readily found by equating the swept up energy, $m_{bo}v_{bo}^2$,
where $m_{bo}=4\pi A R^2_\star R_{bo}$ is 
the swept-up mass, with the energy injected into the shock, $E(v_{bo})$, noting that 
at the breakout radius, $R_{bo}$, the optical depth of the wind,\footnote{
  For simplicity, we adopt the same opacity for the envelope and the wind.}
$\tau_{w,bo} = \kappa m_{bo}/4\pi R_{bo}^2$, satisfies
$\tau_{w,bo}=c/v_{bo}$.  This finally yields
\begin{eqnarray}
v_{bo}=v_0(R_\star/R_{bo})^{2/(\lambda+1)}=v_0(R_\star/v_0t_{bo})^{2/(\lambda+3)}.
\end{eqnarray}
By employing Eqs.~(A2), (A4), and (A7) for $v_0$ in \cite{nakar2010},
the breakout velocity can be expressed in terms of the explosion energy,
$E = 10^{51}E_{51}$ erg, and ejecta mass, $M_{ej}=10 ~M_{10}~M_\odot$, as
\begin{eqnarray}
v_{bo} \simeq 0.1c~E_{51}^{0.44}M_{10}^{-0.31}t_{bo,2}^{-0.25},
\label{eq:v_bo}
\end{eqnarray}
where $t_{bo,2}=t_{bo}/10^2$ s \citep{svirski2014,svirski2014a}.
The above analysis predicts hard X-ray emission with a significant softening 
during the breakout phase for fast shocks, $v_{bo}\simgt 0.1c$, or,
using Eq.~(\ref{eq:v_bo}), for breakouts 
that satisfy $t_{bo,2}\simlt E_{51}^{1.76}M_{10}^{-1.24}$.

\begin{figure}
  \centering
\includegraphics[width=8cm]{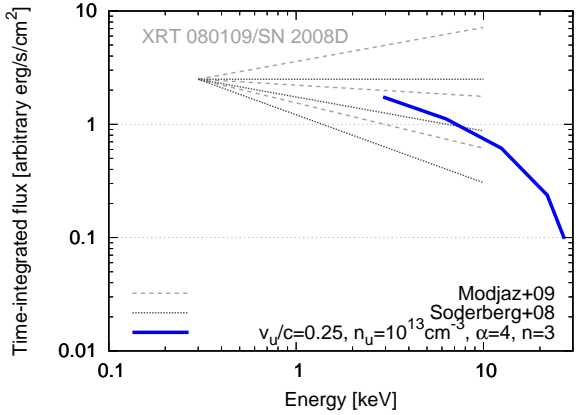}
\caption{Time-integrated spectral energy distribution for
  $v_u/c=0.25, n_u=10^{13}$ cm$^{-3}$, $\alpha=4$, and $n=3$
  taking the temperature softening in Fig.~\ref{fig:Tevo} into account.
  The origin of the vertical axis is arbitrary.
  We also plot the photon spectral indexes of a possible breakout event XRT 080109/SN 2008D,
  $\Gamma=2.1^{+0.3}_{-0.4}$ \citep{Modjaz2009} and $\Gamma=2.3^{+0.3}_{-0.3}$ \citep{Soderberg2008}.
}
\label{fig:spec}
\end{figure}

For illustration, we present in Fig.~\ref{fig:spec} the spectral energy distribution, $\nu F_\nu$, 
integrated over the duration of the breakout phase, assuming that
at any given time the radiation has a thermal spectrum
characterized by the local temperature at the emitting surface
(i.e., the upstream temperature in our solution).
To be precise, the time-integrated spectrum is given formally by
\begin{equation}
\int \nu F_\nu dt \propto \int E(\tau) |f_{\gamma u}(\tau)| \nu^4 \left(e^{h\nu/kT(\tau)}-1\right)^{-1}~d\tau,
\end{equation}
where $\tau\propto r^{-1}\propto v^{\lambda+2}$ is the optical depth at radius $r$ and time $t=r/v$
because the swept-up wind energy $4\pi A r v^2$ is equal to $E(v)\propto v^{-\lambda}$,
$E(\tau)\propto v^{-\lambda}\propto \tau^{-\lambda/(\lambda + 2)}$
is the shock energy at $\tau$, $f_{\gamma u}(\tau)$
is the corresponding escape parameter (see Fig. \ref{fig:tau0-f})
by equating $\tau$ to $\tau_0({\tilde f}_{\gamma u})$ in Eqs.~(\ref{eq:tau0}) and (\ref{eq:alpha}),
and $T(\tau)$ is the temperature (see Fig. \ref{fig:Tevo}).
In Fig.~\ref{fig:spec}, we connect the thermal peaks for corresponding frequencies.
The time-integrated spectrum spreads over a broad frequency range
because of a softening during the breakout phase.

\subsection{Deriving the physical parameters from the observables}\label{sec:test}
\begin{figure}
  \centering
\includegraphics[width=8cm]{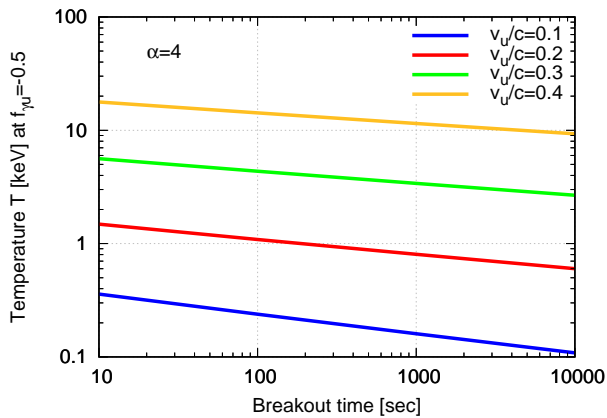}
\caption{Typical temperature at the escape fraction $\tilde{f}_u=-0.5$ with $\alpha=4$
  as a function of breakout time-scale
  for several shock velocities $v_u/c=0.1, 0.2, 0.3$, and $0.4$.
}
\label{fig:T-t}
\end{figure}

The breakout pulse features three general observables --
luminosity, duration, and the typical photon frequency,
which we henceforth associate with the characteristic breakout temperature.
Following the breakout emission, the luminosity declines, sometimes slowly,
as a power law during the transition of the shock
from being radiation dominated to collisionless (see discussion below),
and thus the breakout observables should be associated only
with the rising part of the observed signal, specifically,
the luminosity is roughly the peak luminosity,
the duration is the emission rise time and the temperature is
roughly one-third the peak energy of the time-integrated spectrum of the rising flux.
These three observables are determined by two physical parameters --
the breakout velocity and radius.
Thus, the model is overconstrained, whereby any two observables are
sufficient to determine the physical parameters,
and the third one can be used as a consistency check of the model.
The luminosity and duration are related to the physical parameters
through \citep[e.g.,][]{svirski2014a}:
\begin{eqnarray}
  t_{bo} &\approx& \frac{R_{bo}}{v_{bo}},
  \label{eq:tbo}\\
  L_{bo} &\approx& 1.5 \times 10^{42} t_{bo,2} v_{bo,-1}^3 {\rm ~erg ~s}^{-1}.
  \label{eq:Lbo}
\end{eqnarray}
The dependence of the radiation temperature on the velocity and radius
cannot be expressed as a simple analytic formula (see equation (\ref{eq:analytic}) and Fig.~\ref{fig:Tevo}).
Moreover, as we have seen above it varies with time
(as the escape fraction increases).
However, a rough estimate of the typical breakout temperature
can be obtained by finding the radiation temperature
when the escape fraction is $0.5$.
Figure \ref{fig:T-t} shows the typical temperature for
$\tilde{f}_u=-0.5$ and $\alpha=4$ as a function of breakout time,
for several shock velocities, $v_u/c=0.1, 0.2, 0.3$ and $0.4$.
The density is determined by the opacity condition $n \sigma_T R_{bo}=c/v_{bo}$.
With this Fig.~\ref{fig:T-t}, one can infer the shock velocity
from the observed quantities $T_{bo}$ and $t_{bo}$.
It can then be compared to the velocity obtained from $L_{bo}$,
thereby testing if a given observed flare
may have been the result of a shock breakout from a wind.

\subsection{The observed light curve and spectral evolution}\label{sec:LC}
The above analysis relates the escaping fraction (i.e., luminosity)
and the radiation temperature to 
the instantaneous lab time at which the quasi-steady, planar solution is obtained.
It is naively anticipated 
that this can be directly translated into the observed light curve and
time-resolved spectra during the breakout phase.
However, in practice geometrical (non-planar) effects,
not accounted for by our model, may alter these 
observables (though not the time-integrated spectrum) in ways we now describe.
The main point is that since the breakout takes place once $\tau \approx c/v > 1$,
the photons can still interact with upstream gas also after escaping the shock.
These interactions conceivably include absorption and/or scattering of photons
along their path to the observer.
First, let us consider absorption.
\cite{svirski2012} have shown that when the post-shock plasma is
out of thermodynamic equilibrium then free-free absorption of photons
that escape the shock is negligible.
However, if there are partially ionized metals in the cold upstream
then their X-ray opacity can be considerable.
To estimate the ionization fraction, we use the ionization parameter
$\xi=L/n r^2$, where $L$ is the luminosity of ionizing radiation
and $n$ is the gas density, all expressed in cgs units.
\cite{chevalier2012} (and references therein) have shown that
when $\xi>10^4$, radiation at a temperature of $\sim 10$ keV fully
ionize all metals including iron.
In case of a shock breakout from a stellar wind with $n \propto r^{-2}$,
the ionization depends only on the shock velocity,
roughly as  $\xi \approx 10^8 (v_s/c)^3$.
Consequently, for the shocks considered here ($v_s/c\simgt0.1$)
we have $\xi>10^5$, hence photo-absorption by partially ionized metals is not expected.
Next, let us consider the effect of scattering.
For the shocks we consider here the Thomson optical depth
encountered by a photon escaping the shock is $\sim 2$--$3$.
Thus, a non-negligible fraction of the escaping photons are scattered
at least once on their way to the observer by electrons at radii larger than $R_{bo}$.
This scattering cannot change the emitted spectrum
but it will affect the photon arrival time.
Consequently, our simplified model cannot fully account for the exact shape
of the X-ray light curve nor for the spectral evolution during the rising of the emission. 
Nevertheless, we expect that the emission will show
a hard-to-soft evolution during the rise of the flux,
and the time-integrated spectrum to be non-thermal,
similar to the one shown in Fig.~\ref{fig:spec}. 
  Note that the spectrum is formed in a different way
  from the case of a breakout from a stellar surface,
  in which the time-integrated spectrum is the sum of radiation emitted
  form different positions of the shocked material.

It is worth noting that the breakout emission dominates the rise of the signal and,
perhaps, the initial decay following the peak luminosity 
(over a duration that does not exceed the risetime), but not the entire luminosity evolution. 
The reason is that following the breakout of the RMS a collisionless shock is formed
which is very efficient in converting the shock energy to X-rays \citep{svirski2014},
at least as long as $\tau >1$.
As a result, the observed flux following the breakout phase should exhibit a slow 
power-law decay until the collisionless shock reaches the full extent of the wind \citep{svirski2012}.
The transition from RMS to a fully collisionless shock is anticipated to be gradual, since
the escaping radiation accelerates the plasma ahead of the shock, at radii $r>R_{bo}$,
roughly to a velocity $v \approx v_{bo}(R_{bo}/r)^2$,
and once the shocked gas that trails the radiation, and propagates at a velocity $v_{bo}$
arrives, it drives a collisionless subshock 
that moves at a relative velocity $v_{bo}-v$ into these pre-accelerated fluid.
This subshock strengthens gradually with time until either 
a full conversion is established or until the shock reaches the edge of the wind.
The post-shock electron temperature is set
  by the balance between the heating and (mainly inverse Compton) cooling
  and it is about $\sim 60$ keV and less
  \citep{Katz+11,Murase+11,svirski2012,svirski2014}.
We leave the calculation of the spectrum during this 
phase to a future work.

The only event in which a fast ($>0.1 c$) sub-relativistic shock breakout
from a thick wind was most likely observed is 
the X-ray transient XRT 080109 \citep{Soderberg2008,Mazzali2008,Modjaz2009}.
XRT 080109 is associated with the Type Ibc supernova 2008D,
favoring a WR progenitor, which are known to eject winds
\citep{Tanaka2009,Gal-Yam2014}.
The X-ray peak luminosity is $3.8^{+1}_{-1} \times 10^{43}$ erg s$^{-1}$,
the rise time is $\sim 50$--$100$ s and after the peak
it decays roughly as $t^{-1}$ for about 300 s \citep{Soderberg2008,Modjaz2009}.
The general properties of the signal are in agreement with
the interpretation of a fast shock breakout from a thick wind
\citep{Chevalier2008,svirski2014,svirski2014a}.
In this interpretation, the shock breakout emission dominates
during the rise and at later times,
it is possible that there is also contribution from the collisionless shock
that forms following the breakout.
The radio observations identify synchrotron emission and
the inferred shock radius $\sim 3\times 10^{15}$ cm at $\sim 5$ d
implies a shock velocity of $\sim 0.25$ c \citep{Soderberg2008}.
A simple estimate, $v_{bo} t_{bo}$, implies a breakout radius $R_{bo} \sim 6 \times 10^{11}$ cm
and hence a density $n \sim c/v_{bo} \sigma_T R_{bo} \sim 10^{13}$ cm$^{-3}$.
The observed spectrum is consistent with a power-law spectrum with 
a photon spectral index $\Gamma=2.1^{+0.3}_{-0.4}$ for the period 0--520 s \citep{Modjaz2009}
and $\Gamma=2.3^{+0.3}_{-0.3}$ \citep{Soderberg2008},
showing a significant hard-to-soft evolution \citep{Soderberg2008}.
In Fig.~\ref{fig:spec}, we compare the breakout spectrum computed from the model
with the time-integrated spectral index of XRT 080109,
which includes also emission after the peak and
is therefore most likely not purely the shock breakout spectrum.
It shows that the observations are broadly consistent with the model prediction.
We can also apply the test presented in section \ref{sec:test} to XRT 080109.
Plugging $L_{bo}=4 \times 10^{43}$ erg s$^{-1}$ and $t_{bo}=100$ s into equation (\ref{eq:Lbo})
implies  $v_{bo}/c \approx 0.3$.
Now, using Figure \ref{fig:T-t},
we obtain  $T_{bo} \approx 5$ keV for $v_{bo}/c = 0.3$ and $t_{bo}=100$ s,
in agreement with the observed spectrum.

We should mention that a power-law spectrum could be also produced by other mechanisms.
First, bulk Compton scattering could shape a power-law spectrum
because a typical photon experiences $(c/v)^2$ scatterings,
each one giving a fractional energy increase $\sim (v/c)^2$
and a total average increase of order unity \citep{blandford1981a,blandford1981b,Wang2007,Suzuki2010},
although the efficiency of the non-thermal emission may not be enough \citep{Suzuki2010}.
Second, as above mentioned, the collisionless shock that forms following the breakout,
accelerate electrons to a temperature $\simgt 60$ keV.
These electrons upscatter soft photons to a power-law spectrum \citep{svirski2014,svirski2014a}.
Future observations of shock breakouts that will separate
between the spectrum during the rise and following the peak
will enable us to distinguish between emissions
that comes from the breakout of the RMS
and emission that interacts with electrons accelerated in the collisionless shock that follows.

\section{Conclusions}
We have shown that the breakout of a sub-relativistic, fast ($>0.1 c$) shock
from a thick stellar wind should lead to a broad time-integrated spectrum
during the rise of the observed flux and, conceivably,
softening of the time resolved spectrum,
as shown in Figs.~\ref{fig:Tevo} and \ref{fig:spec}.
The physical reason is that
the photon generation (which decreases the temperature) is enhanced
by the shock steepening during the photon escape.
We applied our results to XRT 080109/SN 2008D and found them
to be consistent with the observed time-integrated,
and the reported softening, of the X-ray spectrum in this source
in Fig.~\ref{fig:spec}.
We also derive a closure relation between the breakout duration,
peak luminosity, and characteristic temperature
in Eqs.~(\ref{eq:tbo}) and (\ref{eq:Lbo}) and Fig.~\ref{fig:T-t},
which is also found to be consistent with the observations of XRT 080109.
Our calculations are based on a semi-analytic model of a planar, RMS that incorporates
photon escape through the upstream plasma, treats radiative transfer
in the diffusion limit, and assumes a quasi-steady
evolution during the breakout phase.

While the time-integrated spectrum of the breakout signal is a robust feature,
the X-ray light curve and the time-resolved spectral evolution 
may be altered by the interaction of the escaping radiation
with the plasma ahead of the RMS, at radii larger than
the breakout radius, where the planar approximation invoked in our analysis breaks down.
In particular, acceleration of the unshocked fluid ahead of the RMS
by the escaping radiation is expected to lead to the gradual
emergence of a collisionless sub-shock that strengthens
over a time comparable to or even longer than the breakout time. 
This intermediate transition from the RMS phase to a fully collisionless shock 
might have interesting observational diagnostics yet to be explored.

\section*{Acknowledgements}

This work was initiated when AL was a visiting professor at
Yukawa Institute for Theoretical Physics (YITP), Kyoto University.
Discussions during the YITP workshop YITP-W-18-11
on 'Jet and Shock Breakouts in Cosmic Transients'
were also useful to complete this work.
This work was partly supported by
Kyoto University, YITP,
the International Research Unit of Advanced Future Studies at Kyoto University,
JSPS KAKENHI nos. 18H01215, 17H06357, 17H06362, 17H06131, and 26287051
(KI),
and the Israel Science Foundation (grant 1114/17) (AL and EN).

\bibliographystyle{mnras}
\bibliography{RMSE}

%

%
%
%

\end{document}